\documentclass[reprint,amsmath,amssymb,aps,prx,balancelastpage,floatfix,superscriptaddress,nofootinbib]{revtex4-2}
\usepackage[T1]{fontenc}
\usepackage{dcolumn}
\usepackage{mathtools}
\usepackage{physics}
\usepackage{times}
\usepackage{eucal}
\usepackage{relsize}
\usepackage{setspace}

\usepackage{graphicx}
\usepackage{bm}
\usepackage{xcolor}
\usepackage{relsize}
\usepackage{hyperref}
\usepackage{hycolor} 

\hypersetup{breaklinks=true,colorlinks=true,citecolor=[rgb]{0,0.18,0.65}, 
urlcolor=[rgb]{0,0.18,0.65}, linkcolor=[rgb]{0,0.18,0.65}, final = true}
\usepackage{bm}
\usepackage{siunitx}
\usepackage[scaled = 0.86]{helvet}

\def\sb#1{\textbf{\textsf{#1}}}
\def\nsb#1{\noindent\sb{#1.}~}

\setcitestyle{round,super}

\newcommand{\eq}{\mathrm{eq}}

\newcommand{\driv}{\mathrm{driv}}
\newcommand{\mx}{\mathrm{max}}
\newcommand{\sym}{\mathrm{sym}}
\newcommand{\ie}{\textit{i.e.,}~}

\begin{document}
\title{\sb{\larger[1.5] Spontaneous Chiral Symmetry Breaking in a Random Driven Chemical System}}

\author{William D. Pi\~{n}eros} 
\affiliation{Center for Soft and Living Matter, Institute for Basic Science (IBS), Ulsan 44919, Korea}
\author{Tsvi Tlusty}
\email{tsvitlusty@gmail.com}
\affiliation{Center for Soft and Living Matter, Institute for Basic Science (IBS), Ulsan 44919, Korea}
\affiliation{Department of Physics, Ulsan National Institute of Science and Technology (UNIST), Ulsan 44919, Korea}
\affiliation{Department of Chemistry, Ulsan National Institute of Science and Technology (UNIST), Ulsan 44919, Korea}

\date{\today}

\begin{abstract}
\textsf{
{\setstretch{1.2}
{\larger[0.5]\textbf{Abstract}\\
Living systems have evolved to efficiently consume available energy sources using an elaborate circuitry of chemical reactions which, puzzlingly, bear a strict restriction to asymmetric chiral configurations. While autocatalysis is known to promote such chiral symmetry breaking, whether a similar phenomenon may also be induced in a more general class of configurable chemical systems---via energy exploitation---is a sensible yet underappreciated possibility. This work examines this question within a model of randomly generated complex chemical networks. We show that chiral symmetry breaking may occur spontaneously and generically by harnessing energy sources from external environmental drives. Key to this transition are intrinsic fluctuations of achiral-to-chiral reactions and tight matching of system configurations to the environmental drives, which together amplify and sustain diverged enantiomer distributions. These asymmetric states emerge through steep energetic transitions from the corresponding symmetric states and sharply cluster as highly-dissipating states. The results thus demonstrate a generic mechanism in which energetic drives may give rise to homochirality in an otherwise totally symmetrical environment, and from an early-life perspective, might emerge as a competitive, energy-harvesting advantage.}}
} 	
\end{abstract}

\maketitle

\section*{\sb{Introduction}}
A hallmark of living systems is their robust and intricate self-organization towards exploiting energy sources from their environment. This can be readily observed in metabolic networks where chemical reactions are carefully orchestrated to transduce external energy sources into molecular energy currency like ATP. Intriguingly, these biochemical reactions take place with a near-exclusive preference for specific chiral configurations over their racemic equivalents, a property termed homochirality.\cite{HomochiralityRev3,HomochiralityReview} 
While the origins of homochirality remain an open question, most mechanisms propose the existence of pre-determinate environmental biases, such as templating-surfaces, polarized light, etc \cite{AminoAcidCrystalTemplate,HomochiralityCrystalTemplate,HomochiralityPolarizedLight}, or more generally rely on autocatalytic features to amplify underlying chiral imbalances\cite{HomoChiralRevAutocatalysisNetworks,HomochiralityAutocatalysisRevBlackmond,HomochiralRev2} which are known to arise in chemical reactions or larger self-assemblying systems.\cite{SoaiReactionRev,HomochiralSelfassemblyBook}
One such classic example is the Frank model in which homochirality emerges spontaneously under the assumption of direct autocatalytic production of a chiral molecule and cross-inhibition with its enantiomer.\cite{FrankOriginal} 
While mutual enantiomer antagonism was thought critical for the model, recent studies by Jafarpour \textit{et al.} showed that stochastic noise in the formation of chiral molecules sufficed to induce the symmetry breaking.\cite{FrankNoise,FrankNoise2} Additionally, a comprehensive study by Laurent \textit{et al}.\cite{GeneralFrankLargeSystems} generalized Frank's single reaction model and demonstrated that chiral symmetry breaking occurred in large, random reaction systems, further corroborating the generality of the result. 

In contrast, whether induced energetic processes, as opposed to strict autocatalysis of elements,\cite{AutocatalysisGeneral} could mediate chiral symmetry breaking events remains an underappreciated but important consideration. This does not imply that the underlying process is necessarily free of cooperative mechanisms, but rather that it depends explicitly on the progression or, feedback from its energy landscape. For instance, energetic coupling through internal or environmental interactions can induce symmetry-breaking transitions, such as the well-known ferromagnetic phase transition. In fact, analogous dipole models have been introduced and successfully applied to explain induced chiral bias in various systems like self-interacting helical polymers and self-assembling nano-crystals.\cite{HelicalFerrochiralModel,NanoCrystalFerroChirality} More broadly, activated energy models like epimerization cycles in polymer models,\cite{AdepOriginal,AdepDetailed} or even achiral particle flow and fluid vortex suspensions are also known to induce spontaneous chiral symmetry breaking.\cite{HydroInteractionsChirality,VertexChirality} Indeed, energetic processes bear natural relevance to the origin of homochirality, as steep non-equilibrium environmental gradients are thought to have conditioned the rise of self-adaptive chemical processes conjectured to predate modern living systems.\cite{OriginLifeRevPhys} 
Thus, whether chiral symmetry breaking could also arise spontaneously in a random chemical system by exploiting environmental energy fluxes seems a natural yet unaddressed question.  

Recently, an example of an energy-driven chemical system was introduced by Horowitz and England and shown to generate highly-dissipating states as a result of continual fuel harvesting from environmental sources~\cite{HorEngChemModel}. The emergence of these states was attributed to dynamical strong matching between system configurations and environmental forcing---a form of feedback---such that the system becomes more responsive to the drive, and thereby enhances its energy fluxes in a reinforcing manner. Similar results have been demonstrated in other driven and self-configurable theoretical and experimental systems.\cite{SelfOrganizedSpringNetwork,OilBeadsElectricField,TubuleSelfAssemblyGTP} These findings then motivate us to ask here whether a similar chiral chemical system could break symmetry via such strong-matched environmental driving. Thus, if such dissipative self-arrangement is sufficiently strong and persistent, then a spontaneous chiral symmetry breaking may occur via amplification of symmetric but inherently noisy conditions. Such result would therefore be entirely distinct from \textit{a priori} autocatalytic assumptions like those of the Frank model, and instead, emerge exclusively out of configurational arrangements conditioned and dependent on the exploitation of an external drive. 
Moreover, an induced dynamic bifurcation by environmental feedback could represent a general symmetry-breaking mechanism by which large interacting networks, like those in evolving living systems, can come to achieve multi-scale, hierarchical organization.\cite{EvolutionaryTransitions,MultiscaleMemory}

In this work, we explore this idea and study a complex, random chiral chemical system subject to configurationally-dependent environmental forces. In particular, the system generates fluctuations of chiral species via random achiral-to-chiral reactions, and the resulting stochastic dynamics is solved in otherwise totally symmetric conditions. Remarkably, we find that for generic but relatively rare instances of our model, chiral symmetry is spontaneously broken, and the transition is entirely dependent on continuous dissipation from the external drives. Moreover, as we elaborate below, this event is tied sharply to the difference in work harnessed along the dynamical trajectory of a chiral-breaking system relative to its symmetric counterpart. Chiral symmetry breaking is therefore predominately observed at high-dissipation regimes. Altogether, this finding demonstrates a first instance where chiral symmetry breaking takes place by virtue of a self-configurable system adapting to an energy source.

\section*{\sb{Results}} 

\begin{figure}[!htb]
\centering
\includegraphics[scale=0.42]{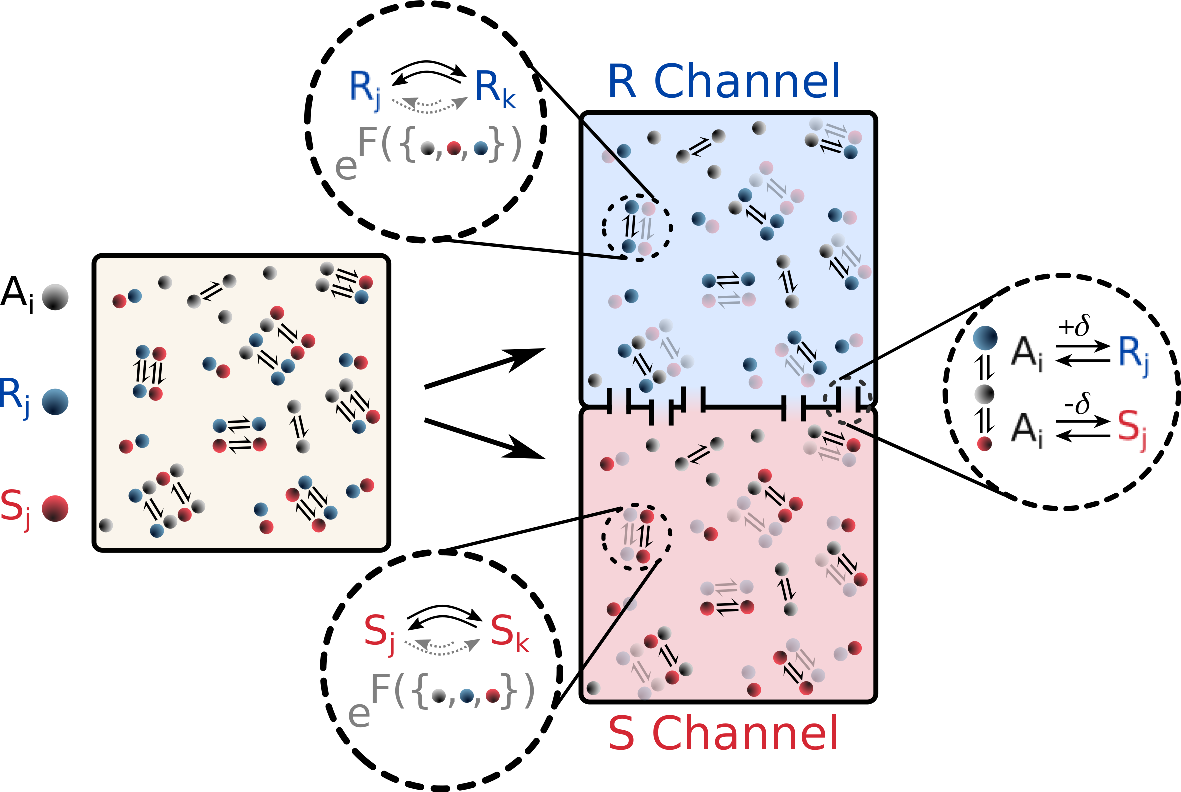} 
\caption{\nsb{Chemical model} Reaction schematic in a solution of achiral elements $A_i$(gray) and chiral elements $R_j(S_j)$, where $R_j$ (blue) is the  chiral counterpart of $S_j$(red). Under total enantioselectivity, the system (yellow box on left) can be decomposed into quasi-independent but mirrored $R$(red) and $S$(blue) configuration channels. Dashed insets show a sample equilibrium chiral reaction between elements $R_j(S_j)$ and $R_k(S_k)$. A fraction of these reactions is randomly doubled and driven by environmental forces $F^\gamma$ which depend on a complex expression of system elements $\{A_i,R_j,S_k\}$. Force expressions are identical but mirrored (\ie $R_j \rightarrow S_j$ and vice-versa) to maintain channel symmetry. The two channels are connected by ``canals'' (rectangular passages) which allow mass exchange between channels via achiral-to-chiral reactions. This exchange is symmetric but may vary by a random fraction $\delta$ as a result of environmental fluctuations.} 
\label{fig:rxnscheme}
\end{figure} 

\nsb{Model: a Complex Chiral Chemical System}
We construct the chemical system using a framework recently introduced in Ref. \citenum{HorEngChemModel}, where achiral species $\{A_i\}$ react under the influence of external environmental forces (for a minimal kinetic picture see supplementary section IX-A). To explore the possibility that coupling to strong environmental drives may induce spontaneous breaking of chiral symmetry, we introduce here additional sets of chiral species $\{R_j\}$ and their corresponding enantiomers $\{S_j\}$. Specifically, we consider a dilute solution of chiral and achiral elements at equilibrium in a container of constant volume and temperature as shown in Figure~\ref{fig:rxnscheme} left. Species are randomly wired in uni- or bi-molecular reactions, some of which may be randomly catalyzed by another independent species. Enantiomers are assumed to be mutually inert, and cross-reaction between chiral species to be perfectly enantioselective. As a result, any chiral species $i$ can be arbitrarily captured by independent $R_i$ and $S_i$ sets since otherwise enantioselectivity with respect to species $j$ would break with a trivial swap of $R/S$ element label. This is illustrated in Figure~\ref{fig:rxnscheme} where the total system may be viewed as consisting of two separate $R/S$ ``channels'' with only the achiral $A_i$ elements held in common. Achiral elements then play the critical role connecting $R/S$ channels as mass exchange ``canals'' through achiral-chiral reactions of the form $A_i \rightleftharpoons R_j(S_j)$ (Figure \ref{fig:rxnscheme}). This exchange process is symmetric on average, but subject to random fluctuations $\pm\delta$ induced by the environment, which manifest as noise in the chiral species dynamics. This small, inherent stochasticity of the system links the quasi-independent channels, thereby allowing for diverging dynamics via strong environmental coupling that would constitute a chirality-breaking event. 

Environmental drives are prescribed here through a set of forces $F=\{F^{\gamma} \}$ that depend on system configurations and induce kinetic tilting of the underlying chemical network. This is achieved by duplicating a random set of existing equilibrium reactions $\alpha$ and favoring (inhibiting) the forward (backward) rate constants by an exponential factor of the applied force, $k^{\prime \pm}_{\alpha}=k^{\pm}_{\alpha} e^{\pm\beta F^{\gamma}/2}$, where $k^{\pm}_{\alpha}$ are equilibrium constants and $\beta=1/k_{\rm B} T$ is the usual inverse temperature and Boltzmann constant. The force itself is a function that may depend on any subset of concentrations, $F^\gamma(\{A_i,R_j,S_k\})$, and whose functional form is constructed from a model known to generate frustrated pair dynamics (see Methods). 
Physically, this forcing form need not imply a detailed accounting of chemical concentrations but may instead be an abstract representation of a system in which chemicals alter their reactivity through changes in their environment. For instance, a photoactivated reaction may lead to precipitates that block light input and inhibit further reaction \ie a state-dependent, environmental effect. 
Indeed, environmental feedback can induce pattern formation and has recently been shown to drive the emergence of phase-separated, active droplets systems.~\cite{ChemomechanicalFeedback,ActiveDroploids}

Lastly, due to the chiral composition of each channel, forces are functionally identical but mirrored, \ie they are related through a corresponding $R/S$ label swap in the expression as $\{A_i,R_k,S_j\} \longleftrightarrow \{A_i,S_k,R_j\}$. This ensures the system is kept symmetric and allows us to deduce the emergence of any diverging dynamics as the result of chiral symmetry breaking and not a prescribed environmental bias.\\

\begin{figure*}[!htb]
\centering
\includegraphics[scale=0.42]{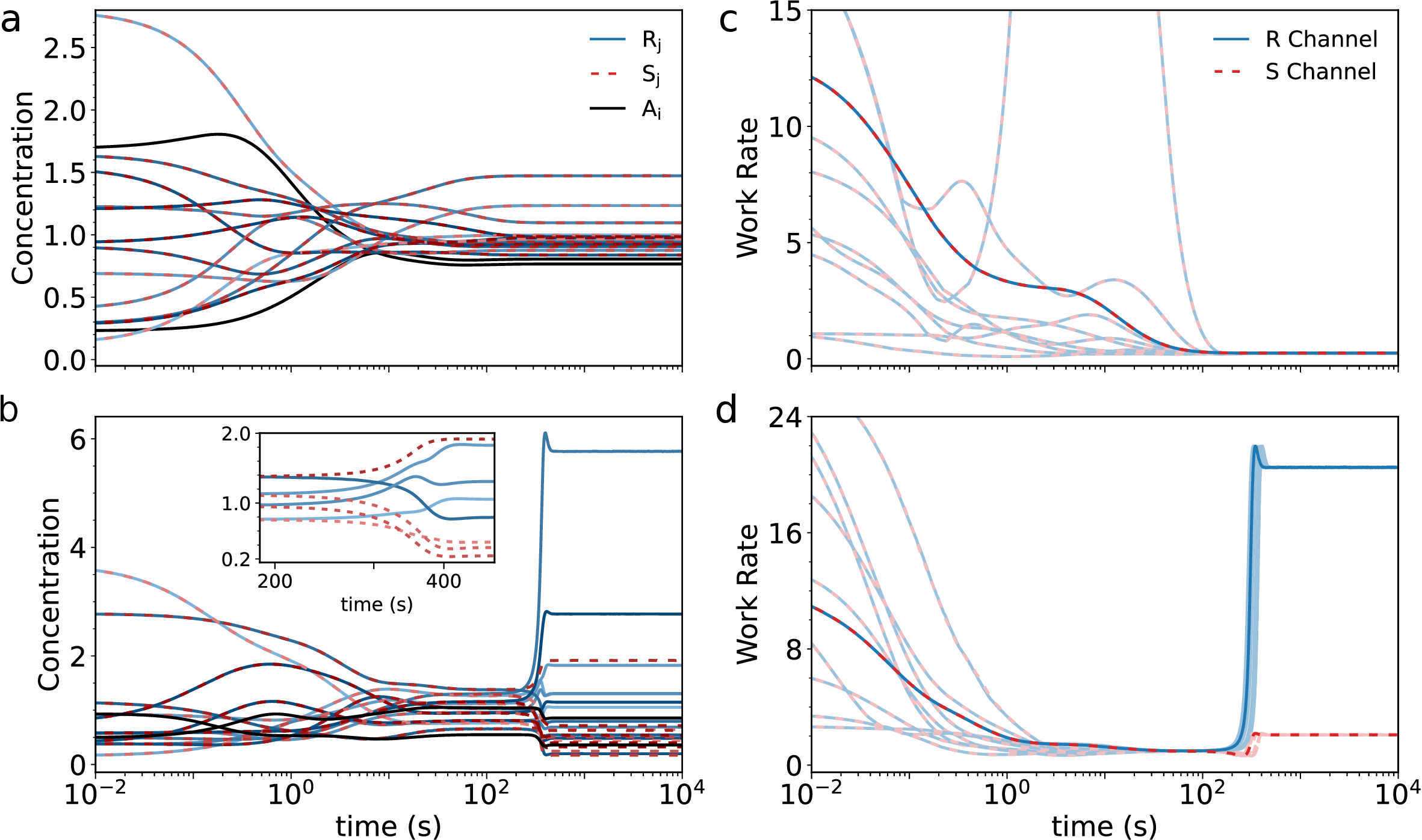} 
\caption{\nsb{Concentrations and work rates of simulated systems} Element concentrations $\{A_i,R_j,S_j\}$ (line shades) as a function of time for a typical chiral-conserving system (a) and a chiral-breaking model (b). Inset shows close-up trajectories of four representative chiral element pairs shortly after chiral breaking, seen here as bifurcations from left to right. Work rate plots as a function of time for the same chiral-conserving (c) and chiral-breaking models (d) for $10$ random initial conditions (lighter lines), including those corresponding to (a) and (b) in darker lines. Here, only $R$ dominant results---defined as the channel achieving the maximum work rate in a given run---are selected and shown for clarity.  
Channel dominance is random and equally distributed on average ($R, S = 50\pm 2\%$, in  $4\times50$ run batches).
}  
\label{fig:concentrations}
\end{figure*} 

\noindent\sb{Highly-dissipating systems may spontaneously break chiral symmetry.} 
~Having defined the chiral chemical model, we now explore our main question: may a totally symmetric system spontaneously break chirality by matching an external environmental drive? To this end, we consider a system of $N_c=12$ chiral pairs and $N_a=2$ achiral elements for a total of $N=26$ elements. The stochastic dynamics of the system are then simulated, starting from random but symmetric initial conditions (\ie $R_i(0) = S_i(0)$, where $R/S(t)$ represent element concentrations as a function of time). Typically, such runs are racemic and replicate the results of Ref. \citenum{HorEngChemModel}, with the majority yielding fixed-point, moderately-driven steady-states and a few displaying highly-dissipating states. Similar results hold for other model parameters (see supplementary Fig. 2). 

In figure~\ref{fig:concentrations}a, we show an example of one such racemic system for a moderately-driven model, where chiral symmetry is clearly conserved across the entire run and elements settle at near-equilibrium concentrations as a result of the moderate drive. To quantify the strength of this driving process, we compute the rate at which work is performed by the drives to power the system, here defined as $\dot{W}=\sum_l F_{l}^{\gamma}|J_{l}|$, where the net rate of formed molecules or reaction currents, $J_l$, is defined as $J_{l}=k^{+}_l X_j-k^{-}_l X_i$. Thus, as seen in figure~\ref{fig:concentrations}c, the overall drive performance is low at the steady-state but applies symmetrically on both chiral channels as per a racemic state. 

In contrast, and very remarkably, for some exceptional instances of our model, the initial symmetrical trajectories of the system can suddenly split and diverge as a result of spontaneous chiral symmetry breaking. This is seen clearly in figure~\ref{fig:concentrations}b for a representative chiral-breaking model, where the concentration profiles of chiral elements bifurcate rapidly at around \SI{200}{s}, and demonstrate strong asymmetric divergence (for comparison of individual elements, see supplementary table 1). Chiral bifurcation is equally apparent in the plots of system work-rates (Figure~\ref{fig:concentrations}d) which in this model manifests as a dominant, highly-driven channel over its weaker chiral counterpart. In addition, as expected from symmetry, the outcome of channel dominance does not depend on its chiral label, but is random and equiprobable upon reruns. Furthermore, the time of symmetry-breaking onset is variable and depends on noise history (see supplementary Figs. 5 and 6). These results hold generically across all obtained chiral-breaking models, suggesting that symmetry breaking is not the result of built-in biases, but originates from the dynamical matching between the reactive system and the environmental drives. \\

\begin{figure*}[!htb]
\centering
\includegraphics[scale=0.58]{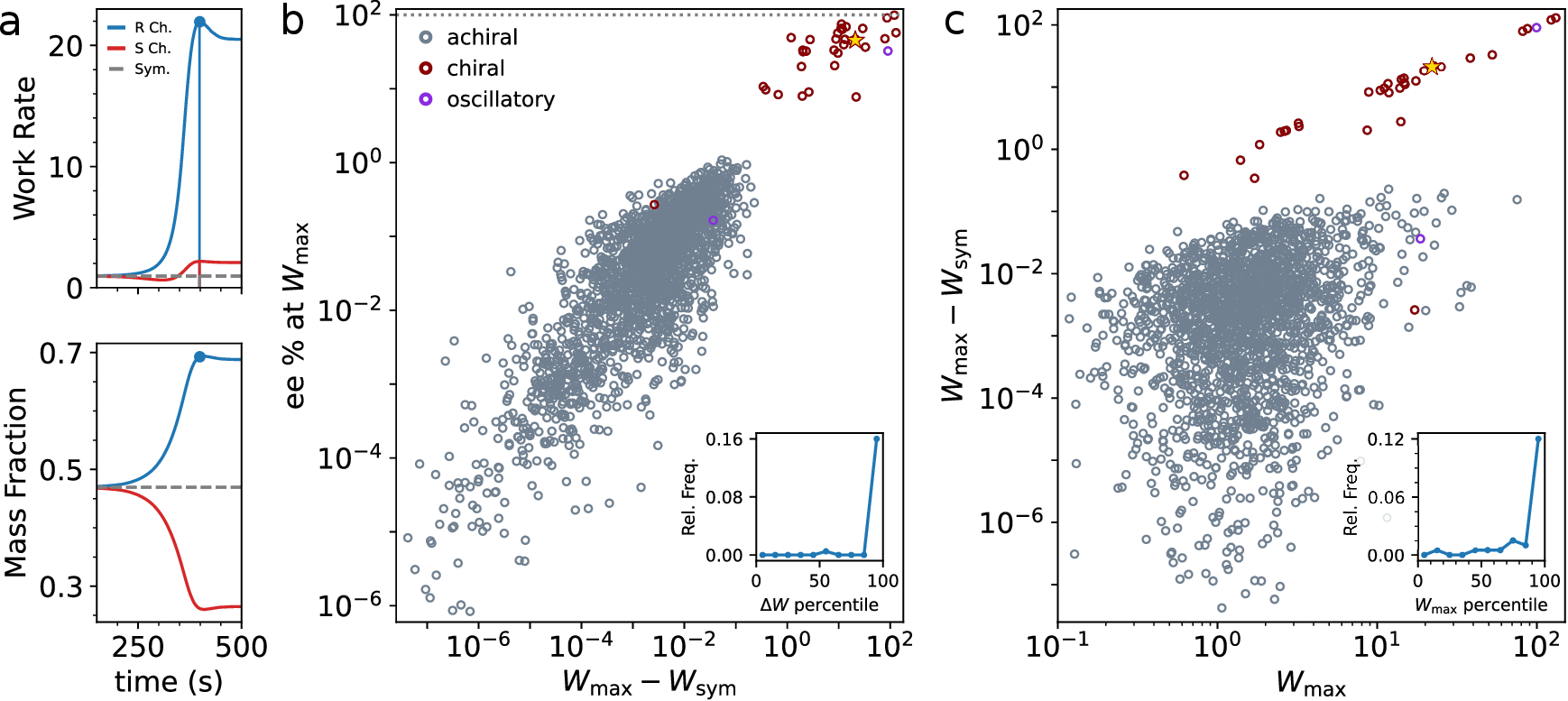} 
\caption{\nsb{Harnessed work and induced asymmetry}
(a) The work rate (top) and mass channel fractions (bottom) for the chiral breaking system in figure~\ref{fig:concentrations} and its corresponding symmetrical case (note linear time axis). Blue marker denotes time point of maximum work rate, and lines under the curve denote the respective work, $W_\mx$, done by the drives over a subsequent $1$ second period. 
(b) The enantiomeric excess $ee$ percent at $W_\mx$ vs. the work difference, $\Delta W = W_\mx-W_\sym$, between the symmetric and chiral breaking trajectories for \num{2000} random, independent models (points below $\sim 10^{-8}$ omitted for clarity). Slate-colored dots denote chiral-conserving models and red dots chiral-breaking models. Yellow star denotes the system drawn in (a). Purple dots denote a very rare class of solutions displaying large $ee$ oscillations as a function of time but with an average $ee=0$ over a complete cycle.  
\textit{Inset}: relative population frequency of chiral-breaking models as a function of the work difference percentile.  
(c) $\Delta W$ vs. $W_\mx$ for the same data set as b. Inset: the relative population frequency of chiral-breaking results as a function of the maximum work percentile.
}
\label{fig:eewmax}
\end{figure*} 
	
\nsb{Large energy differences induce chiral biases} 
Next, we explore this dynamical picture systematically. In particular, we examine how the emergence of chirality-breaking events---which manifest as sudden but reinforcing divergent dynamics---is linked to corresponding sudden energy absorption by the system from the drive. Physically, this could be thought of as an energy injection driving a phase transition. Hence, we identify chiral-breaking models and compare them to perfectly symmetrical counterparts where enantiomers are strictly equal at all times, \ie $S_i(t)=R_i(t)$. Using the same starting conditions, we then re-ran both system versions and computed their difference in maximum harnessed work over an equal-time span after bifurcation, $\Delta W=W_\mx-W_\sym$, where $W_\mx$ and $W_\sym$ represent the work for the actual and symmetric case, respectively. Furthermore, if chiral symmetry breaking depends on $\Delta W$ then it should also correlate with the degree of divergence, here computed as the channel enantiomeric excess, $ee=|m_R-m_S|/(m_R+m_S)$, where $m_{R(S)}$ is the total mass fraction of chiral elements in each channel. As an example, we thus consider again the chiral model in figure~\ref{fig:concentrations}a and plot the work rates (from which $\Delta W$ is derived), and mass fractions $m_{R(S)}$ for the symmetric and chiral system in figure~\ref{fig:eewmax}. Mass bifurcation correlates strongly with elevated work-rate dissipation (hence work performed by the drive) and is consistent with an energy-determined event not present in the symmetrical dynamics. 

The comparison is made systematic by plotting $\Delta W$ against $ee$ for \num{2000} independent models for chiral-breaking and racemic models as shown in Figure~\ref{fig:eewmax}b. Indeed, we find that the values of $ee$ and $\Delta W$ follow a strong positive trend and cluster into distinct sets of chiral-breaking and racemic models, demonstrating that chiral-breaking events are generally followed by large $\Delta W$ differences. This is made evident by computing the frequency of symmetry-breaking systems relative to the population as a function of the percentile ranking of $\Delta W$ as shown in the inset. Symmetry-breaking events are sharply over-represented at exceptionally large values of the harnessed-work difference. Very rarely, $\Delta W$ may be small and still induce chiral symmetry breaking or generate oscillating solutions with large instant $ee$ values but zero average (see supplementary figure 1). As we elaborate later, these rare situations highlight that finite $\Delta W$ differences must also lead to strong matching of the system with the drive, thereby generating and sustaining divergent dynamics.

It is also instructive to examine how $\Delta W$ relates to the overall magnitude of driving extracted by the system as represented by $W_\mx$  (figure~\ref{fig:eewmax}c). Is harnessing a larger overall output from the drive more likely to result in large $\Delta W$ values, for instance? We find that large $\Delta W$ and $W_\mx$ are indeed strongly correlated in the chiral-breaking models and are almost equal since $W_\mx \gg W_\sym$, except for the few outliers at cluster periphery. Likewise, the relative frequency of chiral-breaking models is strongly peaked at high percentiles of $W_\mx$, showing together that strong environmental driving is generally precursive to symmetry breaking. We emphasize that these results hold generically for various instances of the model and the underlying energy landscape features (see supplementary Fig.~2 and Fig.~4).\\

\begin{figure*}[!htb]
\centering
\includegraphics[scale=0.40]{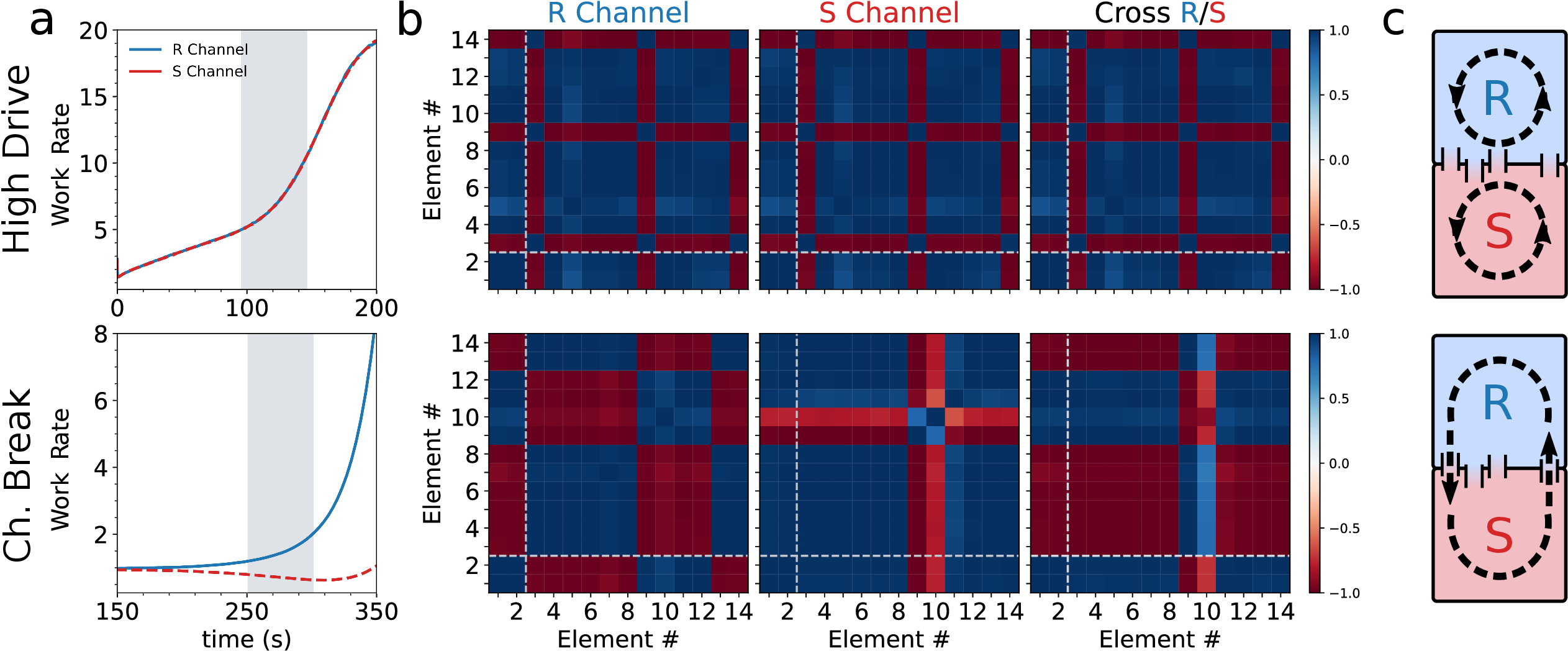} 
\caption{\nsb{System correlations}
(a) Work rates (\ie power) for a high drive system (top) and the chiral breaking model of figure \ref{fig:concentrations} (bottom).
(b) Concentration correlation matrices between all achiral ($1-2$) and chiral ($3-14$) elements of each corresponding $R$ or $S$ channel over a period of $50 s$ (shaded areas in (a)). Dashed light-colored lines denote achiral-chiral element correlation boundaries. Correlation time ranges are chosen to be at comparative timescales of the bifurcation horizon for the chiral-breaking system and the onset of strong-drive in the chiral-conserving model. Third matrix column shows the cross-channel concentration correlation where rows(columns) correspond to the $R$($S$) channel elements. Color bar denotes maximum and minimum values for all matrices, with $+(-)1$ indicating perfect positive(negative) correlation. 
(c) Schematic of channel correlations symbolized by counter-clockwise cycles indicating a strong matching of the system to the environmental drive. For the chiral-conserving case, these are separate and equal (top) but system-wide for the chiral-breaking model (bottom). For a more detailed illustration of the chemical network during this period, see supplementary Figs. 10 and 11.} 
\label{fig:correlations} 
\end{figure*} 

\nsb{System-wide correlations are required to maintain chiral states}
~We now clarify how the character of the system-to-drive coupling determines and sustains different kinds of far-from-equilibrium steady-states. In particular, high dissipating achiral and chiral states are both possible and must clearly differ in mechanism. To examine this issue, we evaluate the correlations coefficients among element concentrations for a representative racemic high-dissipation case and the chiral-breaking system of figure~\ref{fig:concentrations} within and across channels. The correlation is calculated over a time window in which each system is near the onset of strong-driving and for comparative time-scales as shown in figure \ref{fig:concentrations}. Indeed, while both systems display strong intra-channel correlations, the racemic system only replicates correlation patterns across channels as consistent with separate yet identical, mirrored system behavior. On the other hand, channel correlations are distinct for the chiral symmetry breaking case. In particular, elements are strongly anti-correlated across channels, indicative of the sustained but asymmetric system-wide mass transfer taking place. This picture then explains why chiral-breaking states are possible even for relatively small values of $\Delta W$ provided that i) strong correlations with the drive take place system-wide, and ii) they are sustained beyond symmetry breaking onset.

\section*{\sb{Discussion}}
\begin{figure}[!htb]
\centering
\includegraphics[scale=0.38]{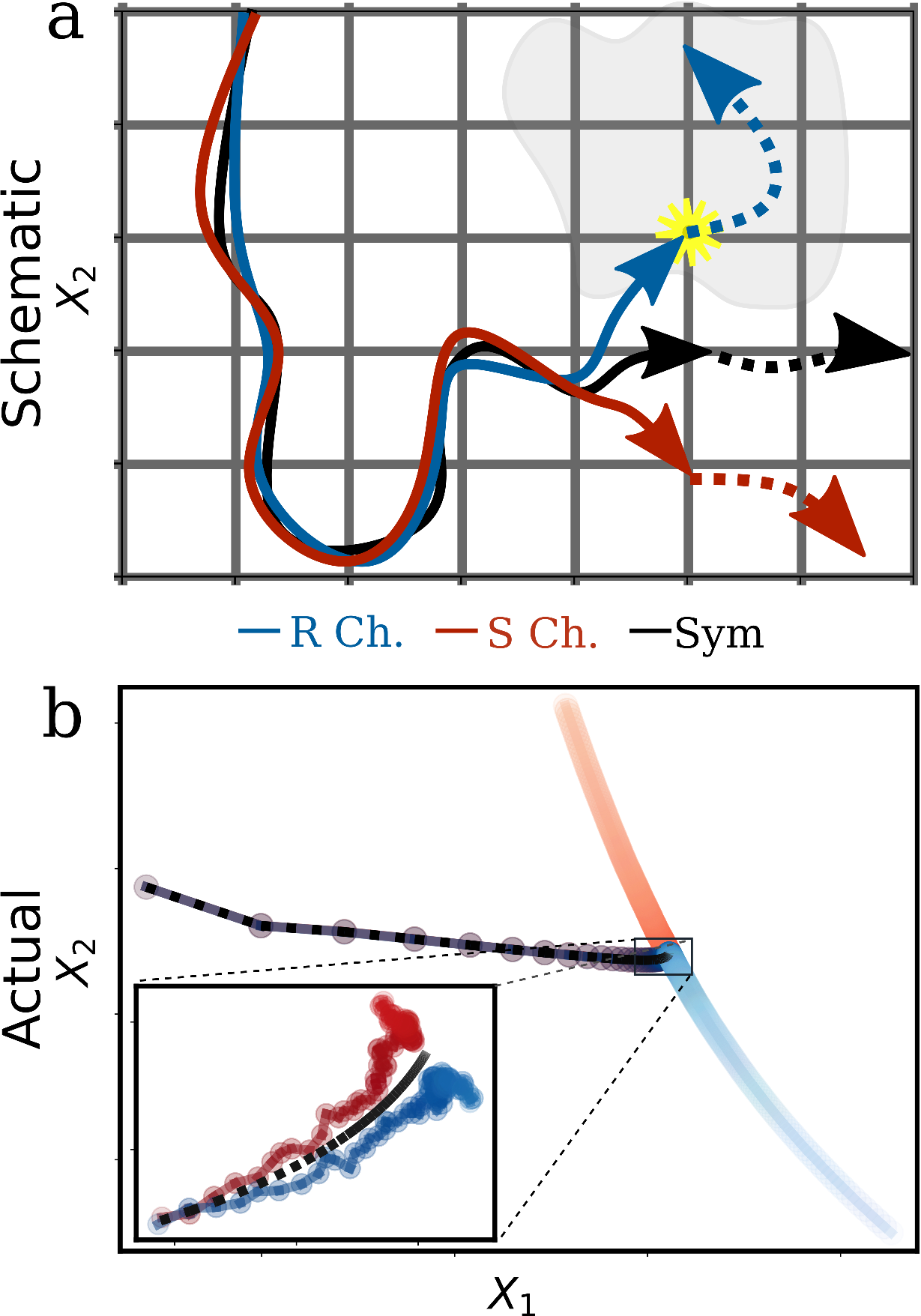} 
\caption{\nsb{Trajectory perspective}
(a) A generalized trajectory perspective of a chirality-breaking event for the $R$(left) and $S$ (right) system channels. Here, $X_1$ and $X_2$ represent arbitrary state-space coordinates, and the shaded area indicates special state-space configurations finely matched to strong driving from the environment.  Yellow asterisk indicates the point where a random fluctuation in the $R$ trajectory (blue) intersects with the special high driving region and triggers a system bifurcation (seen as diverging dashed lines in both $R$ and $S$ trajectories). This intersection event is random and equally likely for both channels upon system restart.
(b) The projected trajectory of the chiral symmetry breaking model of figure \ref{fig:concentrations}, following a singular value decomposition analysis of the corresponding symmetric trajectory. Axes represent the two main eigenvectors (\ie $X_i$) capturing ${\sim}96\%$ of system variation. Time traces are encoded by the respective color from dark to light during a period of $t=2-350$ (s).  
Inset depicts a zoomed time trace near bifurcation onset. 
}  
\label{fig:trajectory}
\end{figure} 

The abrupt change in system dynamics at the chirality transition motivates us to look at the system-drive evolution from a dynamic-trajectory perspective. In particular, all possible system configurations reside in an  $N$-dimensional space of elements' concentrations $X=\{A_i,R_j,S_k\}$. While such a space is admittedly large, one may still envision a projected space of effective system coordinates $X_i$ that capture the system's dominant dynamical modes. Hence, a dynamical trajectory in this space would evolve in $X_i$ until reaching a final steady-state value. 

The complex chiral chemical system of consideration could therefore be pictured as two trajectories representing the $R$ and $S$ channels traversing a parallel but identical state-space as a result of some external drive (Figure~\ref{fig:trajectory}a). Upon introduction of environmental noise in the chiral elements, trajectories may make small deviations from, but still generally follow, the symmetric trajectory. However, for a special class of models, trajectories might graze by strong-driving regions, which, if crossed, will trigger a sudden divergence in channel dynamics. Because trajectory fluctuations are random and generally anti-correlated, only one such channel trajectory will reach this region resulting in the spontaneous breaking of chiral symmetry. 
Naturally, this trajectory crossing-chance is arbitrary but equiprobable upon system restart as expected from the inherent noise symmetry of the reactions. 

Indeed, such schematic picture agrees well with the actual dynamics, as shown in Figure~\ref{fig:trajectory}b. Here, the fully symmetric trajectory is used to perform a singular value decomposition analysis, and its main eigenvectors used to project the unconstrained chiral dynamics over a $t=2-350$ (s) time trace.
Thus, trajectories closely match the symmetric case following initial conditions, but then, somewhere along the visited state space, fluctuations amplify and finally bifurcate. 
This is in contrast to the symmetric trajectory, which, on the other hand, reaches a steady-state point shortly after. Furthermore, as expected, this amplification process coincides with the increased work output by the drives with the main diverging period happening during the period of rapid energy growth (c.f. figure \ref{fig:correlations} a).

This phase space picture then implies that achieving a higher-dissipating state relative to symmetric conditions can induce chiral symmetry breaking and is consistent with the previous analysis of harnessed work. Further, these special regions are only accessible by asymmetric states (through fluctuations) as they exist, by definition, outside of the symmetric state space. As such, sustained symmetry breaking requires and depends exclusively on continual powering from the drives, as otherwise, trajectories would inevitably trace the initial symmetrical state-space, which, by default, conserve racemic conditions. 

We remark that the present results stand in contrast with classic chiral symmetry breaking models where auto-catalysis of one or more chiral elements is the assumed, built-in symmetry breaking mechanism. Instead, here the trajectories diverge as a result of strong matching or positive feedback with the drive, which enhances and reinforces energy production. We emphasize that such an effect does not imply the absence of complex sub-networks. Indeed, looking at the network graphs of such models, we observe many interlocked, dense cyclic patterns at the onset of a strong drive (see supplementary Figs. 10 and 11). However, kinetic rates are not fixed nor prescribed so that the effective network topology is not static but varies dynamically as a function of an induced energy landscape (refer to supplementary section IX-B for further discussion). As such, the observed results represent emergent systems whose special configurations are implicitly conditioned on the exploitation of an environmental source, rather than just \textit{a priori}, built-in features of a network, such as direct autocatalysis of elements. 

As a further example of this distinction, we turn again to the case of a racemic system with high dissipation and a chiral system that exhibits symmetry breaking. In both cases, configurations are strongly matched to their forcings and display an intricate dense network of cycles, yet only one breaks chiral symmetry. Thus, while both systems start and progress as parallel and symmetric $R/S$ channels, only the non-racemic case can reach special asymmetric state-points, which trigger, and then reinforce, a larger energy influx. This implies that despite both racemic and non-racemic cases possessing complex topological elements, it is the induction from an asymmetric input amplified through the drives that powers the symmetry breaking. In this sense, symmetry breaking is not only then confined to built-in features of the network but must arise dynamically and concurrently by feedback from the environmental forcing. 

Interestingly, strong energy dependence with chiral symmetry breaking may be arguably found even in conventional Frank models. For instance, in Ref. \citenum{FrankNoise2}, the degree of homochiral order was shown to correspond to the efficiency of autocatalytic turn-over and hence the strength of dissipated power through supply and consumption of the starting high-energy achiral species.
Likewise, other schematic autocatalytic models also predict the instability of the racemic state past some threshold of energy dissipation \cite{FrankEntropyProduction,RacemateInstabilityEntropyProd}.
In concordance with the findings of this work, these results thus suggest that the emergence of chiral asymmetry could happen simultaneously or shortly after a program of self-sustained energy harvesting has emerged.

More tentatively, the model presented in this work suggests that a chemical network could emerge as some form of a self-stabilizing, dissipative dynamical system through environmental auto-induction by the drives. This scenario could be envisioned as altering of the environment in a manner that reinforces the original reaction network and is therefore ``beneficial" in this sense. For example, a hypothetical set of reactions in a hydrothermal environment could yield products that help dissolve vent-coating minerals and thus boost reagent intake. 
Consequently, while in some circumstances, two equivalent systems may emerge and coexist as highly-dissipative racemic states, for others, a random event might create a small but reinforcing asymmetric ``advantage" which may swiftly amplify and emerge as a single, highly-dissipative asymmetric system.  
Following a generalized notion of ``fitness'' for kinetically-driven chemical systems,\cite{DKstability} one could then possibly say that symmetry breaking is a pathway to achieve such higher ``fitness''.  
The chiral model in this work thus offers an additional mechanism by which a diverse chemical space might establish homochirality, and from an early-life model perspective, may emerge as an inevitable competitive advantage in energy source exploitation.
Dynamically induced bifurcation might then represent a general symmetry-breaking mechanism by which self-configurable networks---much like living systems---learn to efficiently harvest energy, matter and information as they adapt to a continuously changing environment and grow in complexity.

Altogether, we have shown that a complex chemical network---composed of chiral and achiral elements and subject to random environmental drives---may induce spontaneous breaking of chiral symmetry from totally symmetric starting conditions. This process was found to be strongly dependent on the dissipated work difference between the chiral-symmetry-breaking and racemic systems, and is over-represented in systems achieving highly dissipative states. Furthermore, strong, system-wide correlated driving was required to induce chiral-breaking states, as otherwise strong but channel-independent correlations only preserve initial racemic states. 
Strikingly, these results demonstrate that chiral symmetry-breaking can be an induced, energetically-driven process, and stands in analogy to other symmetry-breaking transitions like those in magnetic spin models. 
They also imply, physically, that adaptable chemical systems might exhibit spontaneous chiral symmetry breaking by feedback from a precise energy program---and without initial biases or pre-configured autocatalytic features. 
These observations thus invite us to expand notions of asymmetrical synthesis and pose an intriguing mechanism for the emergence of homochirality in a primordial chemical environment. \\

\section*{\sb{Methods}}
\nsb{Model generation details} 
In this work, we adapt a chemical framework introduced by Horowitz and England \cite{HorEngChemModel} and apply it to a chiral chemical system as follows. 
We consider a well-stirred solution of $N$ chemical elements inside a container of volume $V=1$ and kept at inverse temperature $\beta=1/k_{\rm B}T=1$. 
We then construct a random network of $M$ chemical reactions of the form 
\begin{align}
	\sum_i a^{\alpha}_i X_i \rightleftharpoons \sum_j b^{\alpha}_j X_j,  \quad\quad\quad \alpha=1,\ldots,M, 
\end{align} 
where the $a(b)$ coefficients are either $0$ or $1$ and $X_i \in \{A_i,R_j,S_k\}$ with $N_a$ achiral $A_i$, and $N_c$ chiral $R_j/S_j$ elements. 
Here we assume enantiomers to be mutually inert and perfectly enantioselective with respect to cross-species reactions. 
This is formally equivalent to considering separate $R$ and $S$ element sets and excluding any $R_j/S_k$ reaction pairings. 

Most such reactions are unimolecular, but a fraction may also be bimolecular with probability $p_b$. 
Additionally, catalysts may be added with probability $p_c=0.5$ in the form of an independent element $X_k$ insertion on both sides of the reaction. 
These manifest by modifying rate constants following the mass action kinetics as detailed below.
However, we find that adding catalysts is not entirely necessary though it can enhance the overall yield of non-racemic model solutions (see supplementary Fig. 7).

The rate constants $k^\pm_{\alpha}$ follow mass action kinetics, 
\begin{align}
	k^{+}_{\alpha} = k_{\alpha} \prod_i X_i^{a^{\alpha}_i}~, 
	& \quad\quad
	k^{-}_{\alpha} = k_{\alpha} \prod_j X_j^{b^{\alpha}_i},
\end{align} 
where $k_{\alpha}$ are random base rates chosen uniformly from the set of values $\{\num{e-3},\num{e-2},\num{e-1},1\}$.
A fraction of equilibrium reactions $\eta=M_\mathrm{driven}/M$ are then duplicated, and rates altered by the addition of an environment force as $K^\pm_{\alpha}= k^\pm_{\alpha}e^{\pm\beta F^\gamma/2}$.  
This force form was chosen such that extremal function values from element configurations are rare,\cite{ForceForm,HorEngChemModel} 
and is given as  
\begin{align}
	F^{\gamma}(X) = \sum_{i>j} J^{\gamma}_{ij} (X_i-c^{\gamma}_i) (X_j-c^{\gamma}_j), 	\quad\quad \gamma=1,\ldots,f, 
\end{align} 
where $J_{ij}$ are coupling strengths, of which only  a fraction $\mu$ is non-zero, and chosen randomly from a uniform range between $[-s,s]$.  
The $c^{\gamma}_i$ are random offsets which take a value of either $0$ or $1$.  
Driven reactions are then duplicated and mirrored via an argument swap in the reaction elements (if any) and forces, \textit{e.g.}, $F^\gamma(...A_i R_j ...) \rightarrow F^\gamma(...A_i S_j...)$. This procedure ensures symmetrical driving in the system. 

Finally, system noises are introduced in achiral-to-chiral reactions $A_i \rightleftharpoons R_i(S_i)$ which add (subtract) a fraction $\delta_i$ from $R_i(S_i)$. 
Noises from the equivalent bimolecular reactions are omitted for simplicity and do not alter the nature of the results.  
These fluctuations are modeled as independent white noises for each respective reaction and chosen randomly from a uniform range $[0,\Delta]$, where $\Delta\ll 1$ represents maximum noise strength. 
Thus, each model instance has a unique but systematically generated set of noise terms $\delta_i$ in their dynamics. 

In general, we found chiral-breaking models to arise generically for a wide range of parameters and generally follow the original constraints necessary for high-dissipating systems.\cite{HorEngChemModel} 
Furthermore, we found noise history only altered the onset time of bifurcation but not the steady state in most models, with larger noise leading to shorter onset times (see supplementary Fig. 5). 
As a result, we chose a parameter set that increased the yield of chiral-breaking models per every round of $100$ random generated models. 
For instance, systems with larger proportion of achiral elements yielded lower number of chiral-breaking models due to the larger number of ways mass can transfer between channels and thereby maintain a racemic state (see supplementary Fig. 8).
Larger total system sizes may also be chosen to increase yield (supplementary Fig. 9). 
Final parameters are thus as follows: $N_a=2$, $N_c=2\times12$, $p_b=0.3,\mu=0.25,f=10,s=0.2,\eta=0.6$, and $\Delta=0.02$. \\

\nsb{System simulation} 
The complete system dynamics follow standard mass action formulation but with additional noise terms. 
These noises are considered inherent to a reaction and not replicated for driven reactions. 
The overall chemical kinetics equations are then  

\begin{align} 
\dot{\mathbf{X}}&=(\mathbf{b}_\eq-\mathbf{a}_\eq) (\mathbf{k}^{+}_\eq-\mathbf{k}^{-}_\eq)\cdot(\mathbf{1}+\boldsymbol\delta) \\  \nonumber
				&+(\mathbf{b}_\driv-\mathbf{a}_\driv) (\mathbf{K}^{+}_\driv-\mathbf{K}^{-}_\driv)~, 
\end{align} 
where $\mathbf{k}^\pm_\eq$ and $\mathbf{K}^\pm_\driv$ represent column vectors of the equilibrium, and driven reaction rates respectively, and $\mathbf{b}-\mathbf{a}$ constitutes the corresponding stoichiometric matrix. 
The dot operator is the usual vector product with $\boldsymbol\delta$ representing system noises and $\mathbf{1}$ a unit vector. 
The stochastic differential equations are integrated using an Ito formulation and solved in Mathematica 12 with the ItoProcess method under an Euler-Maruyama scheme. 
To maintain accuracy a time step of $dt=\num{5e-3}$ was used throughout. 
We found runtimes of \num{3e4} steps reached steady-state for most models.\\

\noindent\sb{\larger[0.5]Data availability}\\
\noindent Data supporting the figures within this paper are available from the corresponding authors upon request.\\

\noindent \sb{\larger[0.5]Code availability}\\
The code used for data generation, analytic modeling, and simulations in this study is available from the corresponding authors upon request.\\

\bibliographystyle{naturemag_nourl.bst}
\bibliography{sources}

\begin{thebibliography}{10}
\expandafter\ifx\csname url\endcsname\relax
  \def\url#1{\texttt{#1}}\fi
\expandafter\ifx\csname urlprefix\endcsname\relax\def\urlprefix{URL }\fi
\providecommand{\bibinfo}[2]{#2}
\providecommand{\eprint}[2][]{\url{#2}}

\bibitem{HomochiralityRev3}
\bibinfo{author}{Saito, Y.} \& \bibinfo{author}{Hyuga, H.}
\newblock Colloquium: Homochirality: Symmetry breaking in systems driven far
  from equilibrium.
\newblock \emph{\bibinfo{journal}{Rev. Mod. Phys.}}
  \textbf{\bibinfo{volume}{85}}, \bibinfo{pages}{603--621}
  (\bibinfo{year}{2013}).

\bibitem{HomochiralityReview}
\bibinfo{author}{Blackmond, D.~G.}
\newblock The Origin of Biological Homochirality.
\newblock \emph{\bibinfo{journal}{Cold Spring Harbor Perspectives in Biology}}
  \textbf{\bibinfo{volume}{2}} (\bibinfo{year}{2010}).

\bibitem{AminoAcidCrystalTemplate}
\bibinfo{author}{Hazen, R.~M.}, \bibinfo{author}{Filley, T.~R.} \&
  \bibinfo{author}{Goodfriend, G.~A.}
\newblock Selective adsorption of l- and d-amino acids on calcite: Implications
  for biochemical homochirality.
\newblock \emph{\bibinfo{journal}{Proceedings of the National Academy of
  Sciences}} \textbf{\bibinfo{volume}{98}}, \bibinfo{pages}{5487--5490}
  (\bibinfo{year}{2001}).

\bibitem{HomochiralityCrystalTemplate}
\bibinfo{author}{Weissbuch, I.} \& \bibinfo{author}{Lahav, M.}
\newblock Crystalline Architectures as Templates of Relevance to the Origins of
  Homochirality.
\newblock \emph{\bibinfo{journal}{Chemical Reviews}}
  \textbf{\bibinfo{volume}{111}}, \bibinfo{pages}{3236--3267}
  (\bibinfo{year}{2011}).

\bibitem{HomochiralityPolarizedLight}
\bibinfo{author}{Bailey, J.} \emph{et~al.}
\newblock Circular Polarization in Star- Formation Regions: Implications for
  Biomolecular Homochirality.
\newblock \emph{\bibinfo{journal}{Science}} \textbf{\bibinfo{volume}{281}},
  \bibinfo{pages}{672--674} (\bibinfo{year}{1998}).

\bibitem{HomoChiralRevAutocatalysisNetworks}
\bibinfo{author}{Plasson, R.}, \bibinfo{author}{Kondepudi, D.~K.},
  \bibinfo{author}{Bersini, H.}, \bibinfo{author}{Commeyras, A.} \&
  \bibinfo{author}{Asakura, K.}
\newblock Emergence of homochirality in far-from-equilibrium systems:
  Mechanisms and role in prebiotic chemistry.
\newblock \emph{\bibinfo{journal}{Chirality}} \textbf{\bibinfo{volume}{19}},
  \bibinfo{pages}{589--600} (\bibinfo{year}{2007}).

\bibitem{HomochiralityAutocatalysisRevBlackmond}
\bibinfo{author}{Blackmond, D.~G.}
\newblock Autocatalytic Models for the Origin of Biological Homochirality.
\newblock \emph{\bibinfo{journal}{Chemical Reviews}}
  \textbf{\bibinfo{volume}{120}}, \bibinfo{pages}{4831--4847}
  (\bibinfo{year}{2020}).

\bibitem{HomochiralRev2}
\bibinfo{author}{Ribó, J.~M.}, \bibinfo{author}{Hochberg, D.},
  \bibinfo{author}{Crusats, J.}, \bibinfo{author}{El-Hachemi, Z.} \&
  \bibinfo{author}{Moyano, A.}
\newblock Spontaneous mirror symmetry breaking and origin of biological
  homochirality.
\newblock \emph{\bibinfo{journal}{Journal of The Royal Society Interface}}
  \textbf{\bibinfo{volume}{14}}, \bibinfo{pages}{20170699}
  (\bibinfo{year}{2017}).

\bibitem{SoaiReactionRev}
\bibinfo{author}{Soai, K.}, \bibinfo{author}{Kawasaki, T.} \&
  \bibinfo{author}{Matsumoto, A.}
\newblock Asymmetric Autocatalysis of Pyrimidyl Alkanol and Its Application to
  the Study on the Origin of Homochirality.
\newblock \emph{\bibinfo{journal}{Accounts of Chemical Research}}
  \textbf{\bibinfo{volume}{47}}, \bibinfo{pages}{3643--3654}
  (\bibinfo{year}{2014}).

\bibitem{HomochiralSelfassemblyBook}
\bibinfo{author}{Weissbuch, I.}, \bibinfo{author}{Leiserowitz, L.} \&
  \bibinfo{author}{Lahav, M.}
\newblock \emph{\bibinfo{title}{Stochastic ``Mirror Symmetry Breaking'' via
  Self-Assembly, Reactivityand Amplification of Chirality: Relevance to Abiotic
  Conditions}}, \bibinfo{pages}{123--165} (\bibinfo{publisher}{Springer Berlin
  Heidelberg}, \bibinfo{address}{Berlin, Heidelberg}, \bibinfo{year}{2005}).

\bibitem{FrankOriginal}
\bibinfo{author}{Frank, F.}
\newblock On spontaneous asymmetric synthesis.
\newblock \emph{\bibinfo{journal}{Biochimica et Biophysica Acta}}
  \textbf{\bibinfo{volume}{11}}, \bibinfo{pages}{459--463}
  (\bibinfo{year}{1953}).

\bibitem{FrankNoise}
\bibinfo{author}{Jafarpour, F.}, \bibinfo{author}{Biancalani, T.} \&
  \bibinfo{author}{Goldenfeld, N.}
\newblock Noise-Induced Mechanism for Biological Homochirality of Early Life
  Self-Replicators.
\newblock \emph{\bibinfo{journal}{Phys. Rev. Lett.}}
  \textbf{\bibinfo{volume}{115}}, \bibinfo{pages}{158101}
  (\bibinfo{year}{2015}).

\bibitem{FrankNoise2}
\bibinfo{author}{Jafarpour, F.}, \bibinfo{author}{Biancalani, T.} \&
  \bibinfo{author}{Goldenfeld, N.}
\newblock Noise-induced symmetry breaking far from equilibrium and the
  emergence of biological homochirality.
\newblock \emph{\bibinfo{journal}{Phys. Rev. E}} \textbf{\bibinfo{volume}{95}},
  \bibinfo{pages}{032407} (\bibinfo{year}{2017}).

\bibitem{GeneralFrankLargeSystems}
\bibinfo{author}{Laurent, G.}, \bibinfo{author}{Lacoste, D.} \&
  \bibinfo{author}{Gaspard, P.}
\newblock Emergence of homochirality in large molecular systems.
\newblock \emph{\bibinfo{journal}{Proceedings of the National Academy of
  Sciences}} \textbf{\bibinfo{volume}{118}} (\bibinfo{year}{2021}).

\bibitem{AutocatalysisGeneral}
\bibinfo{author}{Plasson, R.}, \bibinfo{author}{Brandenburg, A.},
  \bibinfo{author}{Jullien, L.} \& \bibinfo{author}{Bersini, H.}
\newblock Autocatalyses.
\newblock \emph{\bibinfo{journal}{The Journal of Physical Chemistry A}}
  \textbf{\bibinfo{volume}{115}}, \bibinfo{pages}{8073--8085}
  (\bibinfo{year}{2011}).

\bibitem{HelicalFerrochiralModel}
\bibinfo{author}{Baumgarten, J.~L.}
\newblock Ferrochirality: A simple theoretical model of interacting dynamically
  invertible helical polymers, 1. The basic effects.
\newblock \emph{\bibinfo{journal}{Macromolecular Rapid Communications}}
  \textbf{\bibinfo{volume}{15}}, \bibinfo{pages}{175--182}
  (\bibinfo{year}{1994}).

\bibitem{NanoCrystalFerroChirality}
\bibinfo{author}{Hananel, U.}, \bibinfo{author}{Ben-Moshe, A.},
  \bibinfo{author}{Diamant, H.} \& \bibinfo{author}{Markovich, G.}
\newblock Spontaneous and directed symmetry breaking in the formation of chiral
  nanocrystals.
\newblock \emph{\bibinfo{journal}{Proceedings of the National Academy of
  Sciences}} \textbf{\bibinfo{volume}{116}}, \bibinfo{pages}{11159--11164}
  (\bibinfo{year}{2019}).

\bibitem{AdepOriginal}
\bibinfo{author}{Plasson, R.}, \bibinfo{author}{Bersini, H.} \&
  \bibinfo{author}{Commeyras, A.}
\newblock Recycling Frank: Spontaneous emergence of homochirality in
  noncatalytic systems.
\newblock \emph{\bibinfo{journal}{Proceedings of the National Academy of
  Sciences}} \textbf{\bibinfo{volume}{101}}, \bibinfo{pages}{16733--16738}
  (\bibinfo{year}{2004}).

\bibitem{AdepDetailed}
\bibinfo{author}{Plasson, R.} \& \bibinfo{author}{Bersini, H.}
\newblock Energetic and Entropic Analysis of Mirror Symmetry Breaking Processes
  in a Recycled Microreversible Chemical System.
\newblock \emph{\bibinfo{journal}{The Journal of Physical Chemistry B}}
  \textbf{\bibinfo{volume}{113}}, \bibinfo{pages}{3477--3490}
  (\bibinfo{year}{2009}).

\bibitem{HydroInteractionsChirality}
\bibinfo{author}{Breier, R.~E.}, \bibinfo{author}{Selinger, R. L.~B.},
  \bibinfo{author}{Ciccotti, G.}, \bibinfo{author}{Herminghaus, S.} \&
  \bibinfo{author}{Mazza, M.~G.}
\newblock Spontaneous chiral symmetry breaking in collective active motion.
\newblock \emph{\bibinfo{journal}{Phys. Rev. E}} \textbf{\bibinfo{volume}{93}},
  \bibinfo{pages}{022410} (\bibinfo{year}{2016}).

\bibitem{VertexChirality}
\bibinfo{author}{S{\l}omka, J.} \& \bibinfo{author}{Dunkel, J.}
\newblock Spontaneous mirror-symmetry breaking induces inverse energy cascade
  in 3D active fluids.
\newblock \emph{\bibinfo{journal}{Proceedings of the National Academy of
  Sciences}} \textbf{\bibinfo{volume}{114}}, \bibinfo{pages}{2119--2124}
  (\bibinfo{year}{2017}).

\bibitem{OriginLifeRevPhys}
\bibinfo{author}{Walker, S.~I.}
\newblock Origins of life: a problem for physics, a key issues review.
\newblock \emph{\bibinfo{journal}{Reports on Progress in Physics}}
  \textbf{\bibinfo{volume}{80}}, \bibinfo{pages}{092601}
  (\bibinfo{year}{2017}).

\bibitem{HorEngChemModel}
\bibinfo{author}{Horowitz, J.~M.} \& \bibinfo{author}{England, J.~L.}
\newblock Spontaneous fine-tuning to environment in many-species chemical
  reaction networks.
\newblock \emph{\bibinfo{journal}{Proceedings of the National Academy of
  Sciences}} \textbf{\bibinfo{volume}{114}}, \bibinfo{pages}{7565--7570}
  (\bibinfo{year}{2017}).

\bibitem{SelfOrganizedSpringNetwork}
\bibinfo{author}{Kachman, T.}, \bibinfo{author}{Owen, J.~A.} \&
  \bibinfo{author}{England, J.~L.}
\newblock Self-Organized Resonance during Search of a Diverse Chemical Space.
\newblock \emph{\bibinfo{journal}{Phys. Rev. Lett.}}
  \textbf{\bibinfo{volume}{119}}, \bibinfo{pages}{038001}
  (\bibinfo{year}{2017}).

\bibitem{OilBeadsElectricField}
\bibinfo{author}{Kondepudi, D.}, \bibinfo{author}{Kay, B.} \&
  \bibinfo{author}{Dixon, J.}
\newblock End-directed evolution and the emergence of energy-seeking behavior
  in a complex system.
\newblock \emph{\bibinfo{journal}{Phys. Rev. E}} \textbf{\bibinfo{volume}{91}},
  \bibinfo{pages}{050902} (\bibinfo{year}{2015}).

\bibitem{TubuleSelfAssemblyGTP}
\bibinfo{author}{te~Brinke, E.} \emph{et~al.}
\newblock Dissipative adaptation in driven self-assembly leading to
  self-dividing fibrils.
\newblock \emph{\bibinfo{journal}{Nature Nanotechnology}}
  \textbf{\bibinfo{volume}{13}}, \bibinfo{pages}{849--855}
  (\bibinfo{year}{2018}).

\bibitem{EvolutionaryTransitions}
\bibinfo{author}{Szathm{\'a}ry, E.} \& \bibinfo{author}{Smith, J.~M.}
\newblock The major evolutionary transitions.
\newblock \emph{\bibinfo{journal}{Nature}} \textbf{\bibinfo{volume}{374}},
  \bibinfo{pages}{227--232} (\bibinfo{year}{1995}).

\bibitem{MultiscaleMemory}
\bibinfo{author}{Libchaber, A.} \& \bibinfo{author}{Tlusty, T.}
\newblock Walking droplets, swimming microbes: on~memory in physics and life.
\newblock \emph{\bibinfo{journal}{Comptes Rendus. M\'ecanique}}
  \textbf{\bibinfo{volume}{348}}, \bibinfo{pages}{545--554}
  (\bibinfo{year}{2020}).

\bibitem{ChemomechanicalFeedback}
\bibinfo{author}{Grinthal, A.} \& \bibinfo{author}{Aizenberg, J.}
\newblock Adaptive all the way down: Building responsive materials from
  hierarchies of chemomechanical feedback.
\newblock \emph{\bibinfo{journal}{Chem. Soc. Rev.}}
  \textbf{\bibinfo{volume}{42}}, \bibinfo{pages}{7072--7085}
  (\bibinfo{year}{2013}).

\bibitem{ActiveDroploids}
\bibinfo{author}{Grauer, J.} \emph{et~al.}
\newblock Active droploids.
\newblock \emph{\bibinfo{journal}{Nature Communications}}
  \textbf{\bibinfo{volume}{12}}, \bibinfo{pages}{6005} (\bibinfo{year}{2021}).

\bibitem{FrankEntropyProduction}
\bibinfo{author}{Kondepudi, D.} \& \bibinfo{author}{Kapcha, L.}
\newblock Entropy production in chiral symmetry breaking transitions.
\newblock \emph{\bibinfo{journal}{Chirality}} \textbf{\bibinfo{volume}{20}},
  \bibinfo{pages}{524--528} (\bibinfo{year}{2008}).

\bibitem{RacemateInstabilityEntropyProd}
\bibinfo{author}{Ribó, J.~M.} \& \bibinfo{author}{Hochberg, D.}
\newblock Spontaneous mirror symmetry breaking: an entropy production survey of
  the racemate instability and the emergence of stable scalemic stationary
  states.
\newblock \emph{\bibinfo{journal}{Phys. Chem. Chem. Phys.}}
  \textbf{\bibinfo{volume}{22}}, \bibinfo{pages}{14013--14025}
  (\bibinfo{year}{2020}).

\bibitem{DKstability}
\bibinfo{author}{Pross, A.}
\newblock Toward a general theory of evolution: Extending Darwinian theory to
  inanimate matter.
\newblock \emph{\bibinfo{journal}{Journal of Systems Chemistry}}
  \textbf{\bibinfo{volume}{2}}, \bibinfo{pages}{1} (\bibinfo{year}{2011}).

\bibitem{ForceForm}
\bibinfo{author}{Sherrington, D.} \& \bibinfo{author}{Kirkpatrick, S.}
\newblock Solvable Model of a Spin-Glass.
\newblock \emph{\bibinfo{journal}{Phys. Rev. Lett.}}
  \textbf{\bibinfo{volume}{35}}, \bibinfo{pages}{1792--1796}
  (\bibinfo{year}{1975}).

\end{thebibliography}

\vspace{0.5cm}
\noindent \sb{\larger[0.5]Acknowledgements}\\
This work was supported by the Institute for Basic Science, Project Code IBS-R020. The authors thank J.M. Horowitz and P. Gaspard for helpful discussions.\\

\noindent \sb{\larger[0.5]Author contributions}\\
\noindent W.D.P. performed the analysis, modeling, and simulations. W.D.P. and T.T. designed the research, wrote and revised the manuscript.\\

\noindent \sb{\larger[0.5]Competing interests}\\
\noindent The authors declare no competing interests.

\end{document}